\documentstyle[12pt]{article}
\begin{document}
\begin{flushright}
\end{flushright}
\hspace{9cm}
\vspace{1cm}
\begin{centering}

{\huge  Hierarchical Wave Functions Revisited }\\
\vspace{1cm}
{ Dingping Li }\\
The Abdus Salam International Center For Theoretical Physics \\
P.O. Box 586, Strada Costiera 11,
Trieste, Italy\\
\end{centering}
\vspace{.25cm}
\begin{abstract}

We study the hierarchical wave functions on a sphere and  on a torus. 
We simplify some wave functions on a sphere or a torus
using the analytic properties of wave functions.   
The open question, the construction of 
the wave function for quasielectron excitations on a torus,  
is also solved in this paper.
\end{abstract}
\vspace{1.25cm}
\begin{flushleft}
ICTP HE 97
\end{flushleft}
{\footnotesize Electronic address: lidp@ictp.trieste.it}


\section{Introduction}

The trial wave function of the fractional quantum Hall effect (FQHE)
on a disk at filling $\nu =1/m$ with $m$ an old integer is given 
by the famous Laughlin wave function \cite{lau}.  
Laughlin wave function had been constructed on a 
spheric or a toric surface later in refs.\ \cite{sphere,torus}.
When $\nu \not= 1/m$, there are several proposals for constructing the
trial wave function,  notably Halperin's  hierarchical wave function
\cite{halp,sphere}  and Jain's composite fermion (CF) wave function
\cite{Jain3}.  In this paper, we will only discuss the 
construction of the hierarchical wave function
(for a review of FQHE, see ref.\ \cite{girvin}).  

Halperin's theory for the FQHE at a hierarchical filling 
is based on the picture of hierarchical 
condensations of quasiparticles. 
The charge and statistics of quasiparticles are
both fractional, and those quasiparticles
were recently observed in experiments \cite{quasi}. 

For example, there are quasiparticle excitations in the case of 
Laughlin's state ($\nu =1/m$). If those quasiparticles
condense and form a Laughlin state them-self, 
a new fractional quantum Hall (FQH) 
state appears with $\nu =1/(m-p)$ or $\nu =1/(m+p)$
($p$ is an even integer), where the minus sign is due to the condensation
of  quasielectrons (QEs),   and the plus sign   is  due to the condensation  of
quasiholes (QHs).  The condensation of quasiparticles at $\nu =1/(m-p)$ or $\nu
=1/(m+p)$ will lead to a higher hierarchical state. The process 
can continue and will form more complicated states. 

Halperin also constructed the wave function at  the hierarchical filling.
The hierarchical wave function had
been also further investigated in refs.\ \cite{Read,wenblok}.  
The  hierarchical wave function  is simply equal to the multiplication of 
those Laughlin wave functions in all hierarchical levels with
all quasiparticle coordinates integrated out (it will become clear in
the next section).  

To construct the hierarchical wave function,
we need to construct the Laughlin wave 
function with quasiparticle excitations.   
The Laughlin wave function  in the presence of 
QH excitations is easy to construct, so does
the hierarchical wave function due to the condensation of QHs
(however the constructed wave function  can not be analytically integrated).  
On the other hand, the Laughlin wave function  
in the presence of  QE excitations is quite complicated and involves derivative operators
in the case of a plane or a sphere. Thus the 
hierarchical wave function due to QE condensations
is quite complicated and intractable in its old form.   
We also note that,  the Laughlin wave function  with  QE excitations 
on a torus is still unknown, and thus the hierarchical 
wave function on a torus due to the condensation of QEs
has not been obtained yet.  We comment that,
although refs.\ \cite{torus2,torus3} had
constructed a hierarchical wave function  at a hierarchical filling on a
torus  due to the condensation of QEs (such wave function on a
sphere was first proposed in ref.\ \cite{Read}),
the constructed wave function  contains  poles (or
singularities) which are difficult to control.

Recently,  some progresses in constructing the hierarchical wave function
had been made in ref.\ \cite{carmem}. In ref.\ \cite{carmem}, it was
discovered that  the Laughlin wave function in the presence of
QEs can be obtained by projecting a rather simple wave function
to the lowest Landau level. Using this fact and the analytic 
properties of the wave functions 
of quasiparticles at all hierarchical levels,  
the hierarchical wave  function  due to the
condensations of QEs can be  greatly simplified
(the wave function can be integrated analytically).
We note that most  of  observed states in experiments
are due to the  condensation of QEs,  only very few 
observed states involves the condensation of QHs.

In this paper, we will generalize   the construction to the case when the
surface is a torus.  How to construct  a Laughlin wave function with
QE excitations on a torus  is an open  problem as noted in
ref.\ \cite{torus}. We will solve this problem by
proposing explicitly a Laughlin wave function   with  QE
excitations on a torus which satisfies all required properties.
Based on this construction,  
a hierarchical wave function due  to the condensation of QEs on a torus can be constructed.  
No singularity exists in the wave function,  
and the  wave function is analytically computable.

In ref.\ \cite{torus1}, the hierarchical wave function 
due to condensations of QHs  was  constructed. 
The wave function of QHs on a torus is  multi-component and the
multi-component  nature of the QH wave function  manifests clearly
in the hierarchical wave function of electrons on a torus.

In refs.\ \cite{torus2,torus3}, a hierarchical wave function on a
torus  due to condensations of QEs is constructed,
although this particular  wave function contains poles. 
In  the contrary to the case of  QHs \cite{torus1},
the  multi-component nature of the wave function of QEs 
does not show in the wave function.  The wave functions of QEs
or QHs must be   multi-component 
as any fractional statistics 
particles on a torus do \cite{ein,eingen}. So it is a paradox.   We will solve  
this paradox in this paper.

We organize the paper as follows:
In the next section, we present a detailed  discussion 
about constructing various 
 hierarchical wave functions on a sphere
(a brief  discussion can be found in ref.\ \cite{carmem}). 
Different hierarchical  wave functions will be constructed  and simplified. 
In section $3$, we will review first what we have known about
the hierarchical wave functions on a torus. Then we will present some
new results, which include solving the puzzling of the 
multicomponent nature of the wave function of QEs
and the construction of a hierarchical wave function 
without any singularity   on a torus due to the condensation of QEs.
Of course in studying the above mentioned problems, the open
problem noted in ref.\ \cite{torus},  
the construction of a Laughlin wave function
with QEs, would be addressed and solved.

\section{Hierarchical wave functions on a sphere}

In this section, we will discuss the construction 
of  the hierarchical wave  function on a sphere  \cite{sphere}.
The quasiparticles satisfy fractional statistics \cite{halp}, and 
the condensation of quasiparticles  will give rise to the FQH state with
\begin{equation}
\nu ={1\over \displaystyle p_1+
{\strut 1\over \displaystyle p_2+
{\strut 1\over \displaystyle \cdots +
{\strut 1\over \displaystyle p_{n}}}}}   ,
\label{filling}
\end{equation}
where $p_1$ is an old positive integer, and $p_i, i\not= 1$
are even integers  (their signs are depend on the types of
the  quasiparticles,  i.e., QH or QE excitations).
  
We use projective coordinates on the sphere (details of notations 
can be found in ref.\ \cite{carmem}). 
The projective  coordinates are given by
$z=2R\cot\frac{\theta}{2} e^{i\varphi}$
and its complex conjugate $\bar z$.  We will take $R=1/2$ for simplicity.

\subsection{Quasiparticle excitations of a Laughlin state}

The Laughlin wave function  at filling $\nu =1/m$ is
\begin{equation}
\Psi_m=\prod_{i<j}^Nd(z_i,z_j)^m,
\end{equation}
where
\begin{equation}
d(z_i,z_j)=\frac{z_i-z_j}
{\sqrt{1+z_i\bar{z_i}}\sqrt{1+z_j\bar{z_j}}}.
\end{equation}
The magnetic flux quanta $\phi$ out of the surface is equal to 
$m(N-1)$ for the Laughlin state.

The Laughlin wave function with the presence 
of   quasiparticle excitations is given by acting
the quasiparticle excitation operators on the original 
Laughlin wave function. 
In the projective coordinates, the operators of the QH excitations 
and the quasielectron excitations are  given in the following form,
\begin{eqnarray}
A^{\dagger}(\omega_k, \bar \omega_k ) \Psi_m(z_i)
& = & \left [ \prod_{j=1}^{N} \prod_{k=1}^{N_q}
d(z_j, \omega_k) \right ] \Psi_m(z_i), \\
A(\omega_k, \bar \omega_k ) \Psi_{m}(z_i) & = & P(\phi )
\left \{  \left [ \prod_{j=1}^{N} \prod_{k=1}^{N_q}
{ d(\bar z_j, \bar \omega_k)} \right ]  \Psi_m(z_i) \right \}
\label{uno}
\end{eqnarray}
where $\omega_k \,\, ,  \bar \omega_k$ are 
the  coordinates of the quasiparticle.  
The flux $\phi$ in the presence of $N_q$ QEs
(QHs) is $m (N-1)-N_q$ ($m(N-1) +N_q$), 
and $P(\phi )$  is an operator which projects a state to 
the lowest Landau level with the magnetic flux quanta equal 
to $\phi$ \cite{carmem}.  Ref.\ \cite{carmem} showed that
the QE excitations given by  eq.\ (\ref{uno}) is 
the same as that in ref.\ \cite{sphere}.

\subsection{Laughlin wave function for quasiparticles}

To construct a hierarchical wave function,
we shall construct  the wave function of quasiparticles.

First we consider a Laughlin state of electrons
and its quasiparticles.
When  the quasiparticles condense, the wave function is 
of Jastrow (or Laughlin) type. The charge of a QH
is $-e/m$ where $e$ is the charge of an electron.  As a QH
has an opposite charge with respect to an electron,
the wave function is anti-holomorphic with the coordinates of QHs.
Because QHs satisfy fractional statistics with 
statistical parameter equal to $1/m$,  the wave function 
for QHs of Laughlin type is found to be:
\begin{equation}
[\Psi_1 (\bar \omega_k )]^{p +1/m},
\end{equation}
where $\Psi_1$ is defined as
\begin{equation}
\Psi_1 (\omega_k )= \prod_{k<l}^{N_q}d(\omega_k, \omega_l). 
\end{equation}
The notation here is  consistent 
with our definition of  the Laughlin wave function.

QEs obey  the same statistics  as QHs.  But as the charge of a QE 
is $e/m$, the wave function of QEs is holomorphic with 
the coordinates of QEs. From these facts, one can deduce that
the wave function of QEs of Laughlin type is 
\begin{equation}
[\Psi_1 ( \omega_k )]^{p - 1/m}
\end{equation}

We have constructed the wave function for quasiparticles  of a Laughlin state 
at filling $\nu =1/m$.   The condensation of those quasiparticles
lead to filling $\nu = 1/(m-1/p)$ or $\nu =1/(m+1/p)$.
Let us define the Laughlin state at $\nu =1/m$ is a  first level
hierarchical state.  The quasiparticles of a Laughlin state can form 
a new Laughlin state of their own, and this will lead a second 
level hierarchical state as just discussed. 
The new Laughlin state then can have their own
quasiparticles. If the new quasiparticles condense, it will lead to 
a third level hierarchical state.  This process can continue, and the
more complicated hierarchical states can be formed. 

Suppose we have a Laughlin state formed by  the quasiparticles
at the $nth$ level
with $\theta_n$ as the statistical parameter. 
Our convention of the definition of $\theta_n$ here is:  
when we exchange the coordinates of two particles, 
we will get the same  phase as we exchange 
two coordinates  of the function  $[\Psi_1 ( z_i^n )] ^{- \theta_n }$.
$\theta_n$ can be a positive or  a  negative rational number, 
but $|\theta_n |$ is generally less or equal than one.  

If the charge of a quasiparticle is negative (we assume that
the charge of an electron is negative), 
the wave function will take the following form,
\begin{equation}
[\Psi_1 ( z_i^n )] ^{p_n - \theta_n }
\label{quasi1}
\end{equation}
where $z_i^n$ is the coordinates of those quasiparticles.
If the charge of the quasiparticle is positive,
then the wave function will take the following form,
\begin{equation}
[\Psi_1 (\bar z_i^n )]^{p_n +\theta_n}.
\label{quasi2}
\end{equation}
$p_n$ is a positive even integer.  

\subsection{The excitations for the Laughlin 
state of quasiparticles}.

In order to  construct  the wave function of the quasiparticles 
at the $n+1th$ level, 
we need to know how to construct the wave function of quasiparticles
of the $nth$ level  with their own quasiparticle excitations.
We shall study the excitation of those states in the previous
subsection. For the QH excitations of  
eq.\ (\ref{quasi1}) and eq.\ (\ref{quasi2}),  the wave functions in the presence of 
those QHs will become 
\begin{equation}
[\Psi_1 ( z_i^n )] ^{p_n - \theta_n } 
\prod_{j=1}^{N_n} \prod_{k=1}^{N_{n+1}} 
d(z_j^n, z_k^{n+1}),
\label{quasin1}
\end{equation}
and 
\begin{equation}
[\Psi_1 (\bar z_i^n )]^{p_n +\theta_n} 
\prod_{j=1}^{N_n} \prod_{k=1}^{N_{n+1}} 
d(\bar z_j^n, \bar z_k^{n+1}) ,
\label{quasin2}
\end{equation}
where $N_n$ is the number of quasiparticles at $nth$ level, 
and $z_i^n$ are the coordinates of those quasiparticles, 
$z_i^{n+1}$ are the coordinates of new quasiparticles 
(we will denote the electron coordinates as $z_i^1$ in this notation)
and $N_{n+1}$ is the number of the new quasiparticles
($N_1$ is the number of electrons).

For the QE excitations of state (\ref{quasi1}) and state (\ref{quasi2}),  
the wave functions will be in the following forms;
\begin{equation}
[\Psi_1 ( z_i^n )] ^{ - \theta_n }           
P (\phi_n -N_{n+1}) \left \{
[\Psi_1 ( z_i^n )] ^{p_n  } 
\prod_{j=1}^{N_n} \prod_{k=1}^{N_{n+1}}
d(\bar z_j^n, \bar z_k^{n+1}) \right \},
\label{quasie1}
\end{equation}
and 
\begin{equation}
[\Psi_1 ( \bar z_i^n )] ^{\theta_n }
P (\phi_n-N_{n+1}) \left \{
[\Psi_1 (\bar z_i^n )]^{p_n }
\prod_{j=1}^{N_n} \prod_{k=1}^{N_{n+1}} 
d( z_j^n,  z_k^{n+1}) \right \},
\label{quasie2}
\end{equation}
where $\phi_n$ is equal to $N_np_n-N_n$, and $P(\phi_n-N_{n+1})$
is an operator which projects the states to the lowest Landau levels
with the magnetic flux $\phi_n-N_{n+1}$ (as previously defined). 
The construction of QEs in eq.\ \ref{quasie1} 
and  eq.\ \ref{quasie2} is a natural generalization of
the construction of  QEs  of the Laughlin state  
(see eq.\ (\ref{uno})).

\subsection{The normalizations of the wave functions}

We need to normalize those states in the previous 
subsection for constructing the  hierarchical wave function.  
The normalized states of  eq.\ (\ref{quasin1}), eq.\ (\ref{quasin2}),
eq.\ (\ref{quasie1}) and eq.\ (\ref{quasie2}) 
will be
\begin{eqnarray}
& & [\Psi_1 ( z_i^n )] ^{p_n - \theta_n }
 [\Psi_1 ( z_i^{n+1} )] ^{1 \over p_n - \theta_n }
\prod_{j=1}^{N_n} \prod_{k=1}^{N_{n+1}} 
d(z_j^n, z_k^{n+1}), \\
& &[\Psi_1 (\bar z_i^n )]^{p_n +\theta_n} 
[\Psi_1 ( \bar z_i^{n+1} )] ^{1 \over p_n + \theta_n }
\prod_{j=1}^{N_n} \prod_{k=1}^{N_{n+1}} 
d(\bar z_j^n, \bar z_k^{n+1}) , \\
& & [\Psi_1 ( z_i^n )] ^{ - \theta_n }           
[\Psi_1 ( z_i^{n+1} )] ^{1 \over p_n - \theta_n }
  \nonumber \\
& & \times P (\phi_n -N_{n+1}) \left \{
[\Psi_1 ( z_i^n )] ^{p_n  } 
\prod_{j=1}^{N_n} \prod_{k=1}^{N_{n+1}}
d(\bar z_j^n, \bar z_k^{n+1}) \right \}, \\
& & [\Psi_1 ( \bar z_i^n )] ^{\theta_n }
[\Psi_1 ( \bar z_i^{n+1} )] ^{1 \over p_n + \theta_n } \nonumber \\
& & \times  P (\phi_n-N_{n+1}) \left \{
[\Psi_1 (\bar z_i^n )]^{p_n }
\prod_{j=1}^{N_n} \prod_{k=1}^{N_{n+1}} 
d( z_j^n,  z_k^{n+1}) \right \}.
\end{eqnarray}
The amplitudes of normalization factors 
can be determined by plasma analogue or 
by the rotational properties of the wave functions, 
and the phase of normalization factors are determined by the statistics of quasiparticles.
The statistical parameters of $\theta_{n+1}$ for the
quasiparticles $z_k^{n+1}$ in  eq.\ (\ref{quasin1}), eq.\ (\ref{quasin2}),
eq.\ (\ref{quasie1}) and eq.\ (\ref{quasie2}) 
are respectively ${1 \over p_n - \theta_n }$, 
$-{1 \over p_n + \theta_n }$, ${1 \over p_n - \theta_n }$, and
$-{1 \over p_n +\theta_n }$. With those parameters, 
we can build the Laughlin states for 
the news quasiparticles with coordinates 
$z_k^{n+1}$ again, thus we can obtain the
hierarchical wave function of any level.

\subsection{Constructions of hierarchical wave functions on a sphere}

We finally come to construct  hierarchical wave functions.
We  first consider when $\nu$ is given by 
\begin{equation}
{1\over \displaystyle p_1-
{\strut 1\over \displaystyle p_2-
{\strut 1\over \displaystyle \cdots -
{\strut 1\over \displaystyle p_{n}}}}}   
\label{filling1}
\end{equation}
with $p_i$ being positive integers. This state is obtained
when the quasiparticles in any level are of QE type. Most of 
observed fillings in experiments can be given by eq.\ (\ref{filling1}).

Take  $n=3$, then the wave function is given as
\begin{equation}
\int 
\prod_{\alpha=1}^{N_2} [dz_{\alpha}^2]
\prod_{\beta=1}^{N_3} [dz_{\beta}^2]
\Psi^1 \Psi^2 \Psi^3,
\end{equation}
where we define that $[dz_{\alpha}^i]$ is equal to 
$\frac{dz_{\alpha}^id\bar z_{\alpha}^i}{
(1+z_{\alpha}^i\bar z_{\alpha}^i)^2}$,  and $\Psi^1$ as 
the normalized wave function of electrons
with quasielectron excitations,
$\Psi^2$ as the  normalized wave function of 
quasiparticles of the electron state, etc.. 
By using formulas previously
obtained  in the last subsection, 
we can write the wave function explicitly, 
\begin{eqnarray}
\Psi = & & \int 
\prod_{\alpha=1}^{N_2} [dz_{\alpha}^2]
\prod_{\beta=1}^{N_3} [dz_{\beta}^3] 
P(p_1N_1-p_1-N_2,  z^1_i)
\nonumber \\ & & 
\times   
\left [ \Psi_1(z_i^1) \right ]^{p_1}
\prod_{i=1}^{N_1} \prod_{\alpha =1}^{N_2} 
d(\bar z_i^1,  \bar z_{\alpha}^2) 
\left [ \Psi_1 (z_i^2) \right ]^{1/p_1} 
\nonumber \\ & & 
\times   
\left [ \Psi_1 (z_i^2) \right ]^{-1/p_1}
P(p_2N_2-p_2-N_3,  z^2_i)
\nonumber \\ & & 
\times   
\left [ \Psi_1 (z_i^2) \right ]^{p_2} 
\prod_{i=1}^{N_2} \prod_{\alpha =1}^{N_3} 
d(\bar z_i^2,  \bar z_{\alpha}^3)  
\left [ \Psi_1 (z_i^3) \right ]^{1 \over p_2 - 1/p_1}
\nonumber \\ & & 
\times   
\left [ \Psi_1 (z_i^3) \right ]^{-1 \over p_2 - 1/p_1}
\left [ \Psi_1 (z_i^3) \right ]^{p_3}, 
\end{eqnarray}
and which can simplify to
\begin{eqnarray}
\Psi = & & \int 
\prod_{\alpha=1}^{N_2} [dz_{\alpha}^2]
\prod_{\beta=1}^{N_3} [dz_{\beta}^3] 
P(p_1N_1-p_1-N_2,  z^1_i)
\nonumber \\ & & 
\times   
\left [ \Psi_1(z_i^1) \right ]^{p_1}
\prod_{i=1}^{N_1} \prod_{\alpha =1}^{N_2} 
d(\bar z_i^1,  \bar z_{\alpha}^2) 
P(p_2N_2-p_2-N_3,  z^2_i)
\nonumber \\ & & 
\times   
\left [ \Psi_1 (z_i^2) \right ]^{p_2} 
\prod_{i=1}^{N_2} \prod_{\alpha =1}^{N_3} 
d(\bar z_i^2,  \bar z_{\alpha}^3)  
\left [ \Psi_1 (z_i^3) \right ]^{p_3}. 
\label{simp}
\end{eqnarray}
We can further simplify the wave function. The key observation 
is that the following equation holds
\begin{equation}
<\psi_1 |P| \psi_2> =<\psi_1 | \psi_2>
\label{key}
\end{equation}
where $\psi_1$ is a state in the lowest Landau level,
and $P$ is an operator which projects $\psi_2$ to the lowest 
Landau Level. In eq.\ (\ref{simp}). 
We consider only the  part of  function which depends
on $z_i^2$,
\begin{eqnarray}
& & \int \prod_{\alpha=1}^{N_2} 
[dz_{\alpha}^2]
\prod_{i=1}^{N_1} \prod_{\alpha =1}^{N_2} 
d(\bar z_i^1,  \bar z_{\alpha}^2) \times
\nonumber \\ & &
P(p_2N_2-p_2-N_3,  z^2_i)
\left [ \Psi_1 (z_i^2) \right ]^{p_2} 
\prod_{i=1}^{N_2} \prod_{\alpha =1}^{N_3} 
d(\bar z_i^2,  \bar z_{\alpha}^3), 
\label{simpl}
\end{eqnarray}
which is equal to the inner product of two bosonic wave functions
as in eq.\ (\ref{key}), when 
\[
\psi_1 =\prod_{i=1}^{N_1} \prod_{\alpha =1}^{N_2} 
d(\bar z_i^1,  \bar z_{\alpha}^2) 
\]
and
\[
\left [ \Psi_1 (z_i^2) \right ]^{p_2} 
\prod_{i=1}^{N_2} \prod_{\alpha =1}^{N_3} 
d(\bar z_i^2,  \bar z_{\alpha}^3).
\]
Thus we can simply omit $P(p_2N_2-p_2-N_3,  z^2_i)$ 
in eq.\ (\ref{simpl}) and  also in eq.\ (\ref{simp}). 
Thus we can write the wave function in a simple form,
\begin{eqnarray}
\Psi = & &  P(\phi ,  z^1_i)
\int
\prod_{\alpha=1}^{N_2} [dz_{\alpha}^2]
\prod_{\beta=1}^{N_3} [dz_{\beta}^3] 
\nonumber \\ & & 
\times   
\left [ \Psi_1(z_i^1) \right ]^{p_1}
\left [ \Psi_1 (z_i^2) \right ]^{p_2} 
\left [ \Psi_1 (z_i^3) \right ]^{p_3}
\nonumber \\ & & 
\times   
 \prod_{i=1}^{N_1} \prod_{\alpha =1}^{N_2} 
d(\bar z_i^1,  \bar z_{\alpha}^2) 
\prod_{i=1}^{N_2} \prod_{\alpha =1}^{N_3} 
d(\bar z_i^2,  \bar z_{\alpha}^3)  
\label{simpkey}
\end{eqnarray}
where $\phi =p_1N_1-p_1-N_2$ is the magnetic flux out of the sphere
(note again that $z^1_i$ are actually the coordinates of electrons).

We can construct the $n$ level hierarchical wave function with
filling given by eq.\ (\ref{filling1}), and then simplify it
following previous discussions about the case of $n=3$.
The wave function is 
\begin{eqnarray}
\Psi = & &  P(\phi ,  z^1_i)
\int \prod_{l=2}^{n} \prod_{k=1}^{N_l} [dz_{k}^l] 
\prod_{l=1}^n \left [ \Psi_1(z_i^l) \right ]^{p_l}
\nonumber \\ & & 
\times  \prod_{l=1}^{n-1} 
\prod_{i=1}^{N_l} \prod_{j =1}^{N_{l+1}} 
d(\bar z_i^l,  \bar z_j^{l+1} ).  
\label{simphi}
\end{eqnarray}
The hierarchical wave function  due to  
the condensations of quasielectrons is amazingly simple,
and it could be calculated analytically. 

Ref.\ \cite{Read} had proposed  a different hierarchical wave function 
with the same filling as the hierarchical wave function
of eq.\ (\ref{simphi}).    The wave function is 
\begin{eqnarray}
\Psi = & &  
\int \prod_{l=2}^{n} \prod_{k=1}^{N_l} [dz_{k}^l] 
\prod_{l=1}^n \left [ \Psi_1(z_i^l) \right ]^{p_l}
\nonumber \\ & & 
\times  \prod_{l=1}^{n-1} 
\prod_{i=1}^{N_l} \prod_{j =1}^{N_{l+1}} 
{1 \over d( z_i^l,   z_j^{l+1} )}.  
\label{ugly}
\end{eqnarray}
However eq.\ (\ref{ugly}) contains singularities and 
we do not know how to control them if we want to
carry out the  calculation of eq.\ (\ref{ugly}). 
If we do the integration in eq.\ (\ref{ugly}),
the wave function will not be holomorphic because of
those singularities, 
thus we need also to project the wave function of
eq.\ (\ref{ugly}) to the lowest Landau level after the
integration.

We could derive eq.\ (\ref{ugly}) by assuming that
the Laughlin wave function 
with QE excitations is 
\begin{equation}
A(\omega_k, \bar \omega_k ) \Psi_{m}(z_i)  = 
\left \{  \left [ \prod_{j=1}^{N} \prod_{k=1}^{N_q}
{1 \over  d( z_j,  \omega_k)} \right ]  \Psi_m(z_i) \right \}
\label{readquasi}
\end{equation}
instead of eq.\ (\ref{uno}), and similarly   eq.\ (\ref{quasie1})
and eq.\ (\ref{quasie2}) are replaced by 
the following equations,
\[
[\Psi_1 ( z_i^n )] ^{ - \theta_n }            
\left \{
[\Psi_1 ( z_i^n )] ^{p_n  } 
\prod_{j=1}^{N_n} \prod_{k=1}^{N_{n+1}}
{1\over d( z_j^n,  z_k^{n+1})} \right \},
\]
and 
\[
[\Psi_1 ( \bar z_i^n )] ^{\theta_n }
\left \{
[\Psi_1 (\bar z_i^n )]^{p_n }
\prod_{j=1}^{N_n} \prod_{k=1}^{N_{n+1}} 
{1\over d( \bar z_j^n,  \bar z_k^{n+1}) }\right \}.
\]
Of course we know that those constructions for QEs
are not favorable as the previous constructions given
by  eq.\ (\ref{uno}),   eq.\ (\ref{quasie1})
and eq.\ (\ref{quasie2}).

All wave functions in the FQHE
on the sphere  shall be  rotationally invariant.
Apply this condition for the wave functions of 
eq.\ (\ref{simphi}) and eq.\ (\ref{ugly}), 
we obtain
\begin{equation}
\left [ \sum_{j=1}^n  \Lambda_{ij}N_j \right ] -\Lambda_{ii}
=\cases {\phi, &if $i=1$; \cr 0, &otherwise, \cr}
\label{4aav}
\end{equation} 
where 
\begin{equation}
\Lambda =\pmatrix{p_1&-1&0&\ldots&0&0\cr
-1&p_2&-1&0&\ldots&0\cr
0&-1&p_3&-1&0&\ldots\cr
\vdots&\vdots&\ddots&\ddots&\vdots&\vdots\cr
\vdots&\vdots&\ddots&\ddots&\vdots&\vdots\cr
0&\ldots&0&-1&p_{n-1}&-1\cr
0&0&\ldots&0&-1&p_n\cr} \, . 
\label{4aat}
\end{equation}

We will now discuss the case when the filling is given by 
eq.\ (\ref{filling}) with $p_i$ all positive.
When $n=3$, the wave function is
\begin{eqnarray}
\Psi = & & \int 
\prod_{\alpha=1}^{N_2} [dz_{\alpha}^2]
\prod_{\beta=1}^{N_3} [dz_{\beta}^3]   
\left [ \Psi_1(z_i^1) \right ]^{p_1}
\prod_{i=1}^{N_1} \prod_{\alpha =1}^{N_2} 
d( z_i^1,   z_{\alpha}^2) 
\left [ \Psi_1 (z_i^2) \right ]^{1/p_1} 
\nonumber \\ & & 
\times   
\left [ \Psi_1 (\bar z_i^2) \right ]^{1/p_1}
\left [ \Psi_1 (\bar z_i^2) \right ]^{p_2} 
\prod_{i=1}^{N_2} \prod_{\alpha =1}^{N_3} 
d(\bar z_i^2,  \bar z_{\alpha}^3)  
\left [ \Psi_1 (\bar z_i^3) \right ]^{1 \over p_2 + 1/p_1}
\nonumber \\ & & 
\times   
\left [ \Psi_1 (z_i^3) \right ]^{1 \over p_2 + 1/p_1}
\left [ \Psi_1 (z_i^3) \right ]^{p_3}, 
\end{eqnarray}
and which can be written as
\begin{eqnarray}
\Psi = & & \int 
\prod_{\alpha=1}^{N_2} [dz_{\alpha}^2]
\prod_{\beta=1}^{N_3} [dz_{\beta}^3] 
\left [ \Psi_1(z_i^1) \right ]^{p_1}
\left [ \Psi_1 (\bar z_i^2) \right ]^{p_2} 
\left [ \Psi_1 (z_i^3) \right ]^{p_3}
\nonumber \\ & & 
\times   
\prod_{i=1}^{N_1} \prod_{\alpha =1}^{N_2} 
d( z_i^1,   z_{\alpha}^2) 
\prod_{i=1}^{N_2} \prod_{\alpha =1}^{N_3} 
d(\bar z_i^2,  \bar z_{\alpha}^3) 
\nonumber \\ & & 
\times   
\left [ \Psi_1 (z_i^2) \Psi_1 (\bar z_i^2)    \right ]^{1/p_1} 
\left [ \Psi_1 (z_i^3) \Psi_1 (\bar z_i^3) \right ]^{1 \over p_2 + 1/p_1} . 
\end{eqnarray}
The hierarchical wave function due to the condensations 
of quasiholes now appears more complicated and it
can not be integrated analytically. 
For any $n$, the wave function is
\begin{eqnarray}
\Psi = & & \int \prod_{l=2}^n
\prod_{\alpha=1}^{N_l} [dz_{\alpha}^l] 
\prod_{ 1\leq 2l-1 \leq n}
\left [ \Psi_1(z_i^{2l-1}) \right ]^{p_{2l-1}}    
\prod_{ 1\leq 2l \leq n}
\left [ \Psi_1(\bar z_i^{2l}) \right ]^{p_{2l}}   
\nonumber \\ & & 
\times    
\prod_{ 1\leq 2l-1 \leq n}
\prod_{i=1}^{N_{2l-1}} \prod_{\alpha =1}^{N_{2l}} 
d( z_i^{2l-1},   z_{\alpha}^{2l})  
\prod_{ 1\leq 2l \leq n}
\prod_{i=1}^{N_{2l}} \prod_{\alpha =1}^{N_{2l+1}} 
d(\bar z_i^{2l},  \bar z_{\alpha}^{2l+1}) 
\nonumber \\ & & 
\times   \prod_{l=2}^n
\left [ \Psi_1 (z_i^l) \Psi_1 (\bar z_i^l)    \right ]^{\epsilon_l}  
\label{wavequasi}
\end{eqnarray}
where $\epsilon_l$ is determined by the relation
\begin{equation}
\epsilon_{l+1}={1\over p_l +\epsilon_l}, \epsilon_l=0.
\end{equation}
In refs.\ \cite{carmem,greiter},  it was 
argued that eq.\ (\ref{wavequasi})
can be approximated by omitting 
$\left [ \Psi_1 (z_i^l) \Psi_1 (\bar z_i^l)    \right ]^{s_l}$
in the integration,
\begin{eqnarray}
\Psi = & & \int \prod_{l=2}^n
\prod_{\alpha=1}^{N_l} [dz_{\alpha}^l] 
\prod_{ 1\leq 2l-1 \leq n}
\left [ \Psi_1(z_i^{2l-1}) \right ]^{p_{2l-1}}    
\prod_{ 1\leq 2l \leq n}
\left [ \Psi_1(\bar z_i^{2l}) \right ]^{p_{2l}}   
\nonumber \\ & & 
\times    
\prod_{ 1\leq 2l-1 \leq n}
\prod_{i=1}^{N_{2l-1}} \prod_{\alpha =1}^{N_{2l}} 
d( z_i^{2l-1},   z_{\alpha}^{2l})  
\prod_{ 1\leq 2l \leq n}
\prod_{i=1}^{N_{2l}} \prod_{\alpha =1}^{N_{2l+1}} 
d(\bar z_i^{2l},  \bar z_{\alpha}^{2l+1}) .
\label{waveomit}
\end{eqnarray}
The wave function of eq.\ (\ref{waveomit}) was also found
to be a good trial wave function comparing the exact
ground state of  a small number of electrons.
Apply the rotational invariance condition on
the wave functions of eq.\ (\ref{wavequasi}) and
eq.\ (\ref{waveomit}),   one finds again 
that eq.\ (\ref{4aav}) must be satisfied, however
the matrix $\Lambda$ is now given by
\begin{equation}
\Lambda =\pmatrix{p_1&+1&0&\ldots&0&0\cr
+1&-p_2&-1&0&\ldots&0\cr
0&-1&p_3&+1&0&\ldots\cr
\vdots&\vdots&\ddots&\ddots&\vdots&\vdots\cr
\vdots&\vdots&\ddots&\ddots&\vdots&\vdots\cr
0&\ldots&0&(-1)^{n-1}&(-1)^n p_{n-1}&(-1)^n\cr
0&0&\ldots&0&(-1)^n&(-1)^{n+1}p_n\cr} \, . 
\label{3aat}
\end{equation}

We have discussed hierarchical wave functions 
due to  the QE condensation or due to 
the QH condensations. Experimentally,  
hierarchical states due to the condensation 
of both QEs and QHs were also observed,
for example, $\nu =n/(4n-1)$ with $n \geq 3$.
$\nu =n/(4n-1)$ can be written as 
\begin{equation}
{1\over \displaystyle p_1+
{\strut 1\over \displaystyle p_2-
{\strut 1\over \displaystyle \cdots -
{\strut 1\over \displaystyle p_{n}}}}}   
\label{filling2}
\end{equation}
with $p_1=3$, $p_i =2, i\not= 1$.
The wave function at $\nu $ of eq.\ (\ref{filling2})
can be constructed straight forwardly. After 
the simplifications, the wave function is 
\begin{eqnarray}
\Psi = & & \left [ \Psi_1( z_i^1) \right ]^{p_1}
\int  \prod_{k=1}^{N_2} [dz_{k}^2]
\prod_{i=1}^{N_1} \prod_{j =1}^{N_{2}} 
d( z_i^1,   z_j^{2} )
\left [ \Psi_1 (z_i^2) \Psi_1 (\bar z_i^2)    \right ]^{1/p_1} 
\nonumber \\ & & 
P(p_2N_2-p_2-N_3 ,  z^2_i)
\int \prod_{l=3}^{n} \prod_{k=1}^{N_l} [dz_{k}^l] 
\prod_{l=2}^n \left [ \Psi_1(\bar z_i^l) \right ]^{p_l}
\nonumber \\ & & 
\times  \prod_{l=2}^{n-1} 
\prod_{i=1}^{N_l} \prod_{j =1}^{N_{l+1}} 
d( z_i^l,   z_j^{l+1} ).  
\label{simpno}
\end{eqnarray}
We could approximate the wave function by taking out
$\left [ \Psi_1 (z_i^2) \Psi_1 (\bar z_i^2)    \right ]^{1/p_1}$
(note that all wave functions discussed in the paper
are trial wave functions, not the exact
wave function), the wave function can be then written as
\begin{eqnarray}
\Psi = & & \left [ \Psi_1( z_i^1) \right ]^{p_1}
\int  \prod_{k=1}^{N_2} [dz_{k}^2]
\prod_{i=1}^{N_1} \prod_{j =1}^{N_{2}} 
d( z_i^1,   z_j^{2} ) \times
\nonumber \\ & & 
\int \prod_{l=3}^{n} \prod_{k=1}^{N_l} [dz_{k}^l] 
\prod_{l=2}^n \left [ \Psi_1(\bar z_i^l) \right ]^{p_l} 
\prod_{l=2}^{n-1} 
\prod_{i=1}^{N_l} \prod_{j =1}^{N_{l+1}} 
d( z_i^l,   z_j^{l+1} ).  
\label{simpmix}
\end{eqnarray}
Now we apply the rotational invariance conditions on the wave functions 
of eq.\ (\ref{simpno}) and eq.\ (\ref{simpmix}).
Again eq.\ (\ref{4aav}) must be satisfied, and 
the matrix $\Lambda$ is now,
\begin{equation}
\Lambda =\pmatrix{p_1&+1&0&\ldots&0&0\cr
+1&-p_2&+1&0&\ldots&0\cr
0&+1&-p_3&+1&0&\ldots\cr
\vdots&\vdots&\ddots&\ddots&\vdots&\vdots\cr
\vdots&\vdots&\ddots&\ddots&\vdots&\vdots\cr
0&\ldots&0&+1&- p_{n-1}&+1\cr
0&0&\ldots&0&+1&-p_n\cr} \, . 
\end{equation}

In this section, we construct various  hierarchical wave functions.
In the case of the hierarchical wave function 
with the filling given by eq.\ (\ref{filling1}),
we greatly simplified  the wave function and obtained 
a very simple formula for the wave function.
In other cases, we can also approximate the wave functions
and  the resulted wave functions also appear in quite simple forms.

\section{Hierarchical wave functions on a torus}

All the wave functions on the sphere constructed 
in the last section can be generalized to the case  when 
the space is a torus. However the construction 
of the similar wave functions on a torus is much more complicated
than  on a sphere.   The wave function of 
a Laughlin wave function on a torus with QE
excitation has been unknown and  the construction of
such wave functions  has remained as  an open question.
We will  solve this open question in this section
and discuss in detail the construction of 
hierarchical wave functions on a torus.

\subsection{Quasiparticles on a torus}

The Laughlin wave function on a torus was obtained in 
ref.\ \cite{torus} 
and was reformulated in refs.\ \cite{torus1,torus2}, 
which we will follow (see also the appendix).

The (normalized) Laughlin wave function of electrons  with
QH excitations on a torus is  
\begin{eqnarray}
\Psi^1(z_i^1, z_\alpha^2)
&=& \exp (-{\pi \phi 
(\sum_i [y_i^1]^2+{1\over p_1} \sum_\alpha [y_\alpha^2 ]^2) \over 
\tau_2})F^1(z_i^1, z_\alpha^2)\, ,\nonumber \\
F^1(z_i^1, z_\alpha^2) & = & \theta {a_1 \brack b_1} 
(\sum_i z_i^1e_1+\sum_\alpha z_\alpha^2 e_1^{\ast}|e_1,\tau)
\prod_{i<j}^{N_1} {[\theta_3(z_i^1-z_j^1|\tau)]}^{p_1} \nonumber \\
& &
\times \prod_{i,\alpha}^{N_1,N_2} [\theta_3(z_i^1-z_{\alpha}^2|\tau)] 
\prod_{\alpha < \beta}^{N_2} 
[\theta_3(z_{\alpha}^2-z_{\beta}^2 |\tau)]^{1\over p_1} 
\, , 
\label{3aauu} 
\end{eqnarray}
where $e_1^2=p_1 \, , e_1^{\ast}={1\over e}$ 
and $a_1, b_1$ are still given by eq.\ (\ref{3aas}), e.g.
$a_1^{\ast}=a_0+l, b_1^{\ast}=b_0$
($a_1,b_1$ are determined by the boundary conditions), 
and magnetic flux $\phi =p_1N_1+N_2$ .

Now we want to construct the QE excitations on a torus. 
Before give the correct construction,
we will construct the  QE  excitations on a torus with singularities 
similar to the  QE excitations on a sphere with singularities
given by eq.\ (\ref{readquasi}):
\begin{eqnarray}
\Psi^1(z_i^1, z_\alpha^2)
&=& \exp (-{\pi \phi 
(\sum_i [y_i^1]^2-{1\over p_1} \sum_\alpha [y_\alpha^2 ]^2) \over 
\tau_2})F^1(z_i^1, z_\alpha^2)\, ,\nonumber \\
F^1(z_i^1, z_\alpha^2) & = & \theta {a_1 \brack b_1} 
(\sum_i z_i^1e_1-\sum_\alpha z_\alpha^2 e_1^{\ast}|e_1,\tau)
\prod_{i<j} {[\theta_3(z_i^1-z_j^1|\tau)]}^{p_1} \nonumber \\
& &
\times \left ( \prod_{i,\alpha} [\theta_3(z_i^1-z_{\alpha}^2|\tau)] 
\prod_{\alpha < \beta} \right )^{-1}
[\theta_3(z_{\alpha}^2-z_{\beta}^2 |\tau)]^{1\over p_1} 
\, , 
\label{3aauue} 
\end{eqnarray}
with $a_1, b_1$ given by eq.\ (\ref{3aas}) again,
and $\phi =p_1N_1-N_2$. 

In order to construct hierarchical wave functions on a torus,
we shall generalize the quasiparticle wave function of
Laughlin type on a sphere to the one on a torus. 

Suppose  that a wave function on a sphere is  
$\prod_{i<j}^N [d(z_i, z_j)]^{p/q}$ with $p$ and $q$ being coprime
to each other.  We find that, on a torus, the wave function 
with $N_q$ QH excitations would be \cite{torus1}:
\begin{eqnarray}
\Psi(z_i, \omega_\alpha)
&=& \exp \left (-{\pi (pN+qN_q)
(p\sum_i [y_i]^2+q \sum_\alpha [y_\alpha(\omega ) ]^2) \over pq
\tau_2} \right )F(z_i, \omega_\alpha   )\, ,\nonumber \\
F(z_i, \omega_\alpha) & = & \theta {a \brack b} 
(\sum_i z_is+\sum_\alpha \omega_\alpha s^{\ast}|e,\tau)
\prod_{i<j} {[\theta_3(z_i-z_j|\tau)]}^{p/q} \nonumber \\
& &
\times  \prod_{i,\alpha} [\theta_3(z_i-\omega_{\alpha}|\tau)] 
\prod_{\alpha < \beta} 
[\theta_3(\omega_{\alpha}-\omega_{\beta} |\tau)]^{q/p} 
\, , 
\label{3aquue} 
\end{eqnarray}
and the wave function with $N_q$ QE excitations (of the type
which contains singularities) :
\begin{eqnarray}
\Psi(z_i, \omega_\alpha)
&=& \exp \left (-{\pi (pN-qN_q)
(p\sum_i [y_i]^2-q \sum_\alpha [y_\alpha(\omega ) ]^2) \over pq
\tau_2} \right )F(z_i, \omega_\alpha   )\, ,\nonumber \\
F(z_i, \omega_\alpha) & = & \theta {a \brack b} 
(\sum_i z_is-\sum_\alpha \omega_\alpha s^{\ast}|e,\tau)
\prod_{i<j} {[\theta_3(z_i-z_j|\tau)]}^{p/q} \nonumber \\
& &
\times  \left ( \prod_{i,\alpha} [\theta_3(z_i-\omega_{\alpha}|\tau)] 
\right )^{-1}
\prod_{\alpha < \beta} 
[\theta_3(\omega_{\alpha}-\omega_{\beta} |\tau)]^{q/p} 
\, , 
\label{3qquue} 
\end{eqnarray}
with $s=(p/q)^{1/2}, s^{\ast}=(q/p)^{1/2}, e=(pq)^{1/2}$.
$a=a^{\ast}e^{\ast},b=b^{\ast}e^{\ast}$  with $e^{\ast}=1/e$
can be determined by the boundary conditions. 
They can be written as (of course there are other ways 
to write out the solutions as we will see later on):
\begin{eqnarray}
& & a^{\ast}=a_0+p\lambda_1 +q\lambda_2 , b^{\ast}=b_0,  \nonumber \\
 & & \lambda_1 =1, \cdots, q,
\lambda_2=1, \cdots , p ,
\end{eqnarray}
with $a_0, b_0$ fixed by the boundary conditions.
We will denote the wave function $\Psi$ as 
$\Psi (z_i, \omega_\alpha | \lambda_1, \lambda_2 )$.
For $z_i$ particles, the wave function has $\lambda_2$ degeneracies
and $\lambda_1$ components 
(the wave function is represented by a column, not a single function). 
For $\omega_i$ particles, the wave function has $\lambda_2$ degeneracies
and $\lambda_1$ components. This fact represents 
the particle-vortex duality in the system.

The wave functions of eq.\ (\ref{3aquue}) and 
eq.\ (\ref{3qquue}) satisfy 
the braid group relations required for 
anyons (particles obeying fractional statistics) on a torus
\cite{ein,eingen}.

We can use eq.\ (\ref{3aquue}) and  eq.\ (\ref{3qquue})
to obtain quasiparticle wave functions on a torus.
Remind that for the wave function with  QE excitations, 
eq. \ (\ref{3qquue}),  the wave function  contains singularities 
and we will return to this point later on.

\subsection{Constructions of 
hierarchical wave functions on a torus}

We will use eq.\ (\ref{3aquue}) and  eq.\ (\ref{3qquue})
to  construct the hierarchical wave function on a torus. 
When $\nu$ is given by eq.\ (\ref{filling}) with
$p_i$ all positive integers,  all quasiparticles are of
QH type,   the wave function was 
obtained in ref.\ \cite{torus1}, and we will not repeat
its derivation here. 

Generally, the wave function of the
$n$ level hierarchy on a torus  can be written as  
\begin{eqnarray}
& & \int \prod_{l=2}^n \prod_{\alpha=1}^{N_l} [dz_{\alpha}^l]
\sum_{\lambda_1, \cdots, 
\lambda_{n-1}} \Psi^1 (z_i^1, z_\alpha^2 |\lambda_1)
 \nonumber \\  & &  \times \cdots
\Psi^l (z_i^l, z_\alpha^{l+1} | \lambda_{l-1}, \lambda_{l}) \cdots
 \times  
\Psi^n (z_i^n | \lambda_{n-1}, \lambda_{n})
\label{tortype}
\end{eqnarray}
with $[dz_{\alpha}^l]=dz_{\alpha}^ld\bar z_{\alpha}^l$. We obtain a
wave function with a degeneracy index  $\lambda_{n}$.

We consider  the case when $\nu$ is given by eq.\ (\ref{filling1}).
First we take the simplest case, $n=2$. Following eq.\ (\ref{tortype}),
the wave function would be:
\begin{eqnarray}
\Psi (z_i^1 )
&=& \exp (-{\pi \phi 
\sum_i [y_i^1]^2 \over 
\tau_2}) \int \prod_{\alpha=1}^{N_2} [dz_{\alpha}^2]
 \prod_{i<j} {[\theta_3(z_i^1-z_j^1|\tau)]}^{p_1}
\nonumber \\ & &  \times  \left ( 
\prod_{i,\alpha} [\theta_3(z_i^1-z_{\alpha}^2|\tau)] \right )^{-1}
\prod_{\alpha < \beta} 
[\theta_3(z_{\alpha}^2-z_{\beta}^2 |\tau)]^{p_2} 
\nonumber \\ & &  \times \sum_{\lambda_1}
\Theta^1 (z_i^1, z_\alpha^2 | \lambda_1) 
\Theta^2 (z_i^2 |\lambda_1, \lambda_2 )
\label{break} 
\end{eqnarray}
where 
\begin{eqnarray}
& & \Theta^1 (z_i^1, z_\alpha^2 | \lambda_1) =\theta {a_1 \brack b_1}
(\sum_i z_i^1s_1-\sum_\alpha z_\alpha^2 s_1^{\ast}|e_1,\tau),
\nonumber \\
& & a_1^{\ast}=a_0 +\lambda_1, \lambda_1=1, \cdots, p_1, b_1^{\ast}=b_0,
\nonumber \\ & &  
s_1=(p_1)^{1/2}, s_1^{\ast}={1\over (p_1)^{1/2}},
e_1=(p_1)^{1/2},
\label{theta1}
\end{eqnarray}
and
\begin{eqnarray}
& & \Theta^2 ( z_\alpha^2 | \lambda_1, \lambda_2 ) 
=\theta {a_2 \brack b_2}
(\sum_\alpha z_\alpha^2 s_2|e_2,\tau),
\nonumber \\
& & a_2^{\ast}=a_0 -\lambda_1 (p_1p_2-1)+ \lambda_2, b^{\ast}_2=b_0
\nonumber \\ & &
\lambda_1=1, \cdots,  p_1, \lambda_2=1, \cdots,  (p_1p_2-1),
\nonumber \\
& & s_2=(p_2- {1\over p_1})^{1/2}, e_2=[p_1(p_1p_2-1)]^{1/2},
\label{theta2}
\end{eqnarray}
and 
\begin{eqnarray}
& & p_1N_1-N_2=\phi, \nonumber \\ 
& & N_1-p_2N_2 =0, 
\label{duereal}
\end{eqnarray}
with $a_0, b_0$ given by eq.\ (\ref{3aas}). 
$a_1, b_1$ are determined by the boundary conditions 
for electrons, $a_2, b_2$ are determined by requiring the 
function within the integration of  eq.\ (\ref{break})
being periodic with coordinates $z_i^2$.  
Those boundary conditions also lead 
to eq.\ (\ref{duereal}).
 
$\lambda_2$ is the degeneracy index
and the degeneracy is thus equal to $p_1p_2-1$,  
agreed with the general results obtained in ref.\ \cite{yongshi}.

We can rewrite 
\begin{equation}
 \sum_{\lambda_1}
\Theta^1 (z_i^1, z_\alpha^2 | \lambda_1) 
\Theta^2 (z_i^2 |\lambda_1, \lambda_2 )
\label{threl}
\end{equation}
as a  theta function on a two dimensional lattice  \cite{mum} .
One can show that a theta function on a two dimensional lattice 
would satisfy the same translational properties 
as the function given by eq.\ (\ref{threl}). The theta function
on a two dimensional lattice is
\[
\theta {a\brack b} (z|e, \tau ) =\sum _{n_i} 
\exp (\pi i {(v+a)}^2 \tau +2\pi i(v+a)\cdot (z+b))\,  ,
\]
with 
\begin{eqnarray}
& & e_i \cdot e_j=A=\pmatrix{p_1&-1\cr
-1&p_2&\cr}, z=\sum_iz_1^2e_1+\sum_{\alpha}z_\alpha^2e_2, 
\nonumber \\
& & a=[{\phi_1 \over 2\pi}+{\Phi +1\over 2}]e^{\ast}_1
+\sum_s n^se^{\ast}_s  \, ,
b=[-{\phi_2 \over 2\pi}+{\Phi +1\over 2}]e^{\ast}_1  \, , 
\nonumber \\ & &
n^se^{\ast}_s \subset {\Lambda^{\ast}\over \Lambda } \, . 
\label{4bbb1}
\end{eqnarray}
$e_i$ defined here as a vector in a lattice
should not be  confused with $e_i$ defined in eq.\ (\ref{theta1})
and eq.\ (\ref{theta2}), we are sorry for the abuse 
of notations.

The linear independent theta functions appeared in eq.\ (\ref{4bbb1})
are given by the independent vectors in 
$n^se^{\ast}_s \subset {\Lambda^{\ast}\over \Lambda }$, 
and the number of such independent vectors is simply 
$\det A =p_1p_2-1$.

Thus we conclude that the linear independent functions 
given by eq.\ (\ref{threl}) can be expressed by 
the linear independent functions given by eq.\ (\ref{4bbb1}).
The discussion of such mathematical relations can be also found
in ref.\ \cite{mum}.  Using this result, the wave function 
of eq.\ (\ref{break}) can be written as
\begin{eqnarray}
\Psi (z_i^1 )
&=& \exp (-{\pi \phi 
\sum_i [y_i^1]^2 \over 
\tau_2}) \int \prod_{\alpha=1}^{N_2} [dz_{\alpha}^2]
 \prod_{i<j} {[\theta_3(z_i^1-z_j^1|\tau)]}^{p_1}
\nonumber \\ & &  \times  \left ( \prod_{i,\alpha} 
[\theta_3(z_i^1-z_{\alpha}^2|\tau)] \right )^{-1}
\prod_{\alpha < \beta} 
[\theta_3(z_{\alpha}^2-z_{\beta}^2 |\tau)]^{p_2} 
\nonumber \\ & &  \times 
\theta {a\brack b} (\sum_i z_i^1e_1+\sum_\alpha z_\alpha^2 e_2  |e, \tau )
\label{breakread} 
\end{eqnarray}
with $a, b, e_i$ given by eq.\ (\ref{4bbb1}).

Remarkably,  the wave function of eq.\ (\ref{breakread})
had been already obtained in refs.\ \cite{torus2,torus3}.
Here we give a derivation of the wave function of eq.\ (\ref{breakread})
based on the picture of the  hierarchical theory.
It was quite puzzling that the wave function of eq.\ (\ref{breakread})
does not show that the wave function of quasiparticles 
is a multicomponent wave function, in contrary to the wave function
obtained in ref.\ \cite{torus1} when $\nu$ is given by eq.\ (\ref{filling}).
Now the puzzle is solved because the wave function 
of eq.\ (\ref{breakread}) is equivalent to the wave function of
eq.\ (\ref{break}),  and eq.\ (\ref{break}) 
shows that the wave function of quasiparticles is a multicomponent wave function. 
 
We can repeat the discussion for an arbitrary $n$, generalizing 
the wave function of eq.\ (\ref{break}) to the one for
the $n$ level hierarchy, and then simplifying it by
using the  relations of   
the theta function on a  $n$ dimensional lattice,
we will get:
\begin{eqnarray}
\Psi (z_i^1 )
&=& \exp (-{\pi \phi 
\sum_i [y_i^1]^2 \over 
\tau_2}) \int \prod_{l=2}^n \prod_{\alpha=1}^{N_l} [dz_{\alpha}^l]
\nonumber \\ & &
\prod_{l=1}^n \prod_{i<j} {[\theta_3(z_i^l-z_j^l|\tau)]}^{p_l}
\prod_{l=1}^{n-1}
\left ( \prod_{i,\alpha}^{N_l, N_{l+1}} 
[\theta_3(z_i^l-z_{\alpha}^{l+1}|\tau)] \right )^{-1}
\nonumber \\ & &  \times 
\theta {a\brack b} 
(\sum_{i,l} z_i^le_l |e, \tau )
\label{breakread1} 
\end{eqnarray}
with 
\begin{eqnarray}
& & e_i \cdot e_j=A=\pmatrix{p_1&-1&0&\ldots&0&0\cr
-1&p_2&-1&0&\ldots&0\cr
0&-1&p_3&-1&0&\ldots\cr
\vdots&\vdots&\ddots&\ddots&\vdots&\vdots\cr
\vdots&\vdots&\ddots&\ddots&\vdots&\vdots\cr
0&\ldots&0&-1&p_{l-1}&-1\cr
0&0&\ldots&0&-1&p_l\cr} \, ,
\nonumber \\
& & a=[{\phi_1 \over 2\pi}+{\Phi +1\over 2}]e^{\ast}_1
+\sum_s n^se^{\ast}_s  \, ,
b=[-{\phi_2 \over 2\pi}+{\Phi +1\over 2}]e^{\ast}_1  \, , 
\nonumber \\ & &
n^se^{\ast}_s \subset {\Lambda^{\ast}\over \Lambda } \, ,
\nonumber \\ & &
A_{ij}N_j=\phi \delta_{i,1}.
\label{4bbb2}
\end{eqnarray}
Because the wave function contains singularities, 
the wave function does not represent a state on
the lowest Landau level and one needs to project
the wave function to the lowest Landau level in the end.
The degeneracy is given the number of independent vectors
in $n^se^{\ast}_s \subset {\Lambda^{\ast}\over \Lambda }$
and it is equal to $\det A$.

\subsection{Quasielectrons on a torus}

We note that the QE wave function used in our
construction of the wave function of eq. (\ref{breakread1})
contains singularities. 
We will construct a Laughlin wave 
function with QE 
excitations on a torus without singularities, which
was mentioned as  an open problem  in ref.\ (\cite{torus})
and still is  an open problem today. 

The Laughlin wave function with QE excitations on a sphere can
be written as a derivative  operator  acts on the Laughlin 
wave function \cite{sphere}. It is very difficult to
generalize this derivative operator on a sphere 
to an operators on a torus as suggested in ref.\ \cite{torus}.  However 
the Laughlin wave function with QE excitations on a sphere can
be also written as the projection to the lowest Landau level 
by a wave function which contains  
higher Landau level states  as described in eq.\ (\ref{uno}).  
We will find that it is actually 
quite {\bf easy} to generalize
eq.\ (\ref{uno}) to the case on a torus. What we need to do is
to replace 
\[
\left ( \prod_{i,\alpha} [\theta_3(z_i-\omega_{\alpha}|\tau)] 
\right )^{-1}
\]
in  eq.\ (\ref{3aauu}) by a function,  
which is regular,  satisfies the same translational properties
(which must hold for any trial wave function),
and  is not necessary a holomorphic function. 

The important observation is that the function 
\[
{\overline {\theta_3 (z|\tau)}} \exp \left [
{\pi (z-\bar z)^2 \over 2\tau_2} \right ]
\]
with ${\overline {\theta_3 (z|\tau)}}$ as the complex
conjugate of $ \theta_3 (z|\tau)$ has the same translational 
properties as
\[
{1\over \theta_3 (z|\tau)}.
\]
Thus we propose that the Laughlin wave function with QE excitations
on a torus is 
\begin{eqnarray}
\Psi^1(z_i^1, z_\alpha^2)
&=& P(\phi, z_i^1) \exp (-{\pi \phi 
(\sum_i [y_i^1]^2-{1\over p_1} \sum_\alpha [y_\alpha^2 ]^2) \over 
\tau_2})F^1(z_i^1, z_\alpha^2)\, ,\nonumber \\
F^1(z_i^1, z_\alpha^2) & = & \theta {a_1 \brack b_1} 
(\sum_i z_i^1e_1-\sum_\alpha z_\alpha^2 e_1^{\ast}|e_1,\tau)
\prod_{i<j} {[\theta_3(z_i^1-z_j^1|\tau)]}^{p_1} \nonumber \\
& &
\times  \prod_{i,\alpha} 
{\overline {\theta_3(z_i^1-z_{\alpha}^2|\tau)}}
\exp \left [
{\pi (z_i^1-z_{\alpha}^2-\bar z_i^1+\bar z_{\alpha}^2)^2 
\over 2\tau_2} \right ]
\nonumber  \\ & & 
\times \prod_{\alpha < \beta} 
[\theta_3(z_{\alpha}^2-z_{\beta}^2 |\tau)]^{1\over p_1} 
\, , 
\label{qereal} 
\end{eqnarray}
where $P(\phi, z_i^1)$ with $\phi =p_1N_1-N_2$
is an operator which projects the wave function
\begin{equation}
\exp (-{\pi \phi 
(\sum_i [y_i^1]^2-{1\over p_1} \sum_\alpha [y_\alpha^2 ]^2) \over 
\tau_2})F^1(z_i^1, z_\alpha^2)
\label{anareal}
\end{equation}
to the lowest Landau level. All parameters in eq.\ (\ref{qereal})
are the same as those in eq.\ (\ref{3aauue}).

One can check the wave function
of eq.\ (\ref{anareal}) satisfies the translational properties 
required for the electrons on a torus 
with magnetic flux $\phi =p_1N_1-N_2$ (see the appendix). 
However the state represented by the wave function of
eq.\ (\ref{anareal}) does not lie purely
on the lowest Landau level and one needs to project it to the 
lowest Landau level in the end to obtain the 
Laughlin wave function with QE excitations on a torus. 
We note that eq.\ (\ref{qereal}) is a direct generalization 
of eq.\ (\ref{uno}) which is a construction 
of the similar wave function  on a sphere. 
In the case of eq.\ (\ref{uno}), one can replace the projection operator and
function $d(\bar z_j, \bar \omega_k) $ by a derivative operator.
However in the case of eq.\ (\ref{qereal}), one can {\bf not} find a
similar derivative operator for constructing QE excitations
on a torus as  in the case of a sphere  (at least we do not 
know how to do that now).  It seems that the construction of
QE excitations involves intrinsically higher Landau levels on a torus.
The construction on the torus can also 
be generalized to the  case when the surface is a   high 
genus Riemann surface or  other complicated surfaces 
(the Laughlin wave function on a Riemann surface 
was obtained in ref.\ (\cite{iengoli}).
We finally comment that it seems that 
the construction of the Laughlin wave function 
with QE excitations given by eq.\ (\ref{qereal}) is quite unique
and we do not know any other plausible constructions. 
Such construction by eq.\ (\ref{qereal}) could (and should)
be checked by numerical calculations.

We can also construct QE excitations for the Laughlin wave 
function of quasiparticles. We will not discuss it here,
and we comment that it involves  more deep understandings of 
the singular  gauge for anyons on a torus
(see ref.\ \cite{iengo} for the detailed discussions 
of the singular gauge on a torus).

\subsection{Hierarchical wave functions on a torus revisited}

We can now construct the hierarchical wave function 
on a torus by using the construction of QE excitations
discussed in the previous subsection.
The part of the wave function which is dependent on the 
center coordinates is  the same as  the one we 
obtained by using the construction of QE excitations
with singularities. We have a trick to derive
the wave function without doing 
similar calculations done in the previous subsections.    
The trick is that 
we simply replace all functions like 
$(\theta_3(z))^{-1}$ by $\overline {\theta_3(z)} 
\exp { \pi (z-\bar z)^2 \over 2\tau_2}$
in the wave function of  eq.\ (\ref{breakread1}), and we get
\begin{eqnarray}
\Psi (z_i^1 )
&=& P(\phi, z_i^1)\exp (-{\pi \phi 
\sum_i [y_i^1]^2 \over 
\tau_2}) \int \prod_{l=2}^n \prod_{\alpha=1}^{N_l} [dz_{\alpha}^l]
\nonumber \\ & & \times
\prod_{l=1}^n \prod_{i<j} {[\theta_3(z_i^l-z_j^l|\tau)]}^{p_l}
\prod_{l=1}^{n-1}
 \prod_{i,\alpha}^{N_l, N_{l+1}} 
{\overline {\theta_3(z_i^l-z_{\alpha}^{l+1}|\tau)} }
\nonumber \\ & & \times
\exp { {\pi [z_i^l-z_{\alpha}^{l+1}-\bar z_i^l+\bar z_{\alpha}^{l+1}]^2}
\over 2\tau_2} 
\theta {a\brack b} 
(\sum_{i,l} z_i^le_l |e, \tau ),
\label{breakread3} 
\end{eqnarray}
with all parameters given by eq.\ (\ref{4bbb2}).

Now we show an example when $\nu $ is given by eq.\ (\ref{filling})
(detailed discussions can be found in ref.\ \cite{torus1}).
The wave function at $\nu ={1\over p_1 +1/p_2}$  is  
\[
\int \prod_{\alpha=1}^{N_2} [dz_{\alpha}^2] 
\sum_{\lambda_1} 
\Psi^1(z_i^1, z_\alpha^2|\lambda_1) 
\Psi^2( z_\alpha^2|\lambda_1, \lambda_2)
\]
where $\Psi^1(z_i^1, z_\alpha^2|\lambda_1) $ 
is given by  eq.\ (\ref{3aauu}) with $a^{\ast}=a_0+\lambda_1$, 
and $\Psi^2( z_\alpha^2|\lambda_1, \lambda_2)$ 
is given by the following equation,
\begin{eqnarray}
& & \Psi^2 (z_{\alpha}^2|\lambda_1, \lambda_2) 
= \exp [-{  
\pi \phi   \sum_\alpha (y_{\alpha}^2)^2 \over 
p_1 \tau_2}]
F^2(z_{\alpha}^2 |\lambda_1, \lambda_2) 
\, ,\nonumber \\ & &
{\overline {F^2(z_{\alpha}^2 |\lambda_1, \lambda_2)}} 
= \theta {a_2\brack b_2} 
(\sum_\alpha z_{\alpha}^2 s_2|e_2, \tau) 
\prod_{\alpha < \beta} [\theta_3(z_{\alpha}^2-z_{\beta}^2 |\tau)]^
{{1\over p_1}+p_2}  \, ,  
\label{3bba} 
\end{eqnarray}
where 
\begin{eqnarray}
& & e_2=[p_1(p_1p_2+1)]^{1/2} \, , s_2=[p_2+{1\over p_1}]^{1/2}
\nonumber \\ & &
a_2=a_2^{\ast}e_2, b_2=b_2^{\ast}, 
\nonumber \\ & &
a_2^{\ast}= a_0+ \lambda_1(p_1p_2+1)+ \lambda_2p_1, 
b_2^{\ast}=b_0
\nonumber \\ & &
\lambda_1 =1, \cdots , p_1, \lambda_2 =1, \cdots , p_1p_2+1,
\nonumber \\ & &
p_1N_1+N_2=\phi \, , N_1-p_2N_2=0 .
\label{theta5}
\end{eqnarray}
Thus the wave function is
\begin{eqnarray}
\Psi (z_i^1 )
&=& \int  \prod_{\alpha=1}^{N_2} [dz_{\alpha}^2]
 \exp  \left (-{\pi \phi 
(\sum_i [y_i^1]^2+{2\over p_1} \sum_\alpha [y_\alpha^2 ]^2) \over 
\tau_2} \right ) \nonumber \\ & & \times
\prod_{i<j}^{N_1} [\theta_3(z_i^1-z_j^1|\tau)]^{p_1} 
\prod_{\alpha< \beta}^{N_2} \left [
{\overline {\theta_3(z_{\alpha}^2-z_{\beta}^2|\tau)} }\right ]^{p_2}
\nonumber \\ & &
\times \prod_{i,\alpha}^{N_1,N_2} [\theta_3(z_i^1-z_{\alpha}^2|\tau)] 
\prod_{\alpha < \beta}^{N_2} 
|\theta_3(z_{\alpha}^2-z_{\beta}^2 |\tau)|^{2\over p_1} 
\nonumber \\ & & 
\times \sum_{\lambda_1}
\Theta^1 (z_i^1, z_\alpha^2 | \lambda_1) 
\Theta^2 (z_i^2 |\lambda_1, \lambda_2 ),
\label{mixing}
\end{eqnarray}
where 
\[
\Theta^1 (z_i^1, z_\alpha^2 | \lambda_1) 
=\theta {a_1(\lambda_1) \brack b_1} 
(\sum_i z_i^1e_1+\sum_\alpha z_\alpha^2 e_1^{\ast}|e_1,\tau).
\]
with all parameters are given by eq.\ (\ref{theta1}), 
except that $a_1^{\ast}=a_0+\lambda_1$, and
\[
\Theta^2 (z_i^2 |\lambda_1, \lambda_2 )
=\overline {\theta {a_2(\lambda_1, \lambda_2)
\brack b_2} (\sum_\alpha z_{\alpha}^2 s_2|e_2, \tau)}
\]
where $a_2(\lambda_1, \lambda_2)$ is given by
eq.\ (\ref{theta5}).
We note that  
\[
\sum_{\lambda_1}
\Theta^1 (z_i^1, z_\alpha^2 | \lambda_1) 
\Theta^2 (z_i^2 |\lambda_1, \lambda_2 )
\] in eq.\ (\ref{mixing})
is not equal to a theta function on a two dimensional lattice as
in the case of $\nu ={1 \over p_1 -1/p_2}$, because 
$\Theta^2 (z_i^2 |\lambda_1, \lambda_2 )$ is now an anti-holomorphic
function. 

As  in the case of the corresponding wave function on a sphere, 
the wave function of eq.\ (\ref{mixing}) also can not be integrated
out analytically. Similarly as discussed in the previous section,
we can approximate eq.\ (\ref{mixing})
by a wave function which is analytically integrable.
We can replace 
\[
\prod_{\alpha < \beta}^{N_2} 
|\theta_3(z_{\alpha}^2-z_{\beta}^2 |\tau)|^{2\over p_1} 
\]
in eq.\ (\ref{mixing})
by 
\[
\exp \left [ -{\pi \over 2p_1 \tau_2}
\sum_{\alpha < \beta}^{N_2}  
(z_{\alpha}^2-z_{\beta}^2- \bar z_{\alpha}^2+ \bar z_{\beta}^2)^2 
\right ].
\]

\section{Conclusions}

We have discussed various hierarchical wave functions on a sphere
and on a torus. The wave functions can be simplified
by using the analytical properties of the wave functions.
We also gave a derivation of the hierarchical wave function 
due to the condensations of QEs on a torus proposed in  
ref.\ \cite{torus2,torus3}. The wave function 
for quasiparticles must be  multicomponent 
even in the case of QEs.  We also solved an open problem, 
the construction of  the Laughlin wave function
with QE excitations on a torus.

\section{Acknowledgments}

We would like to thank ICTP for financial supports.
We also thank ISI Foundation of Torino for
giving us the opportunity to present partial results
of this paper in the workshop of 
``Quantum Hall Effect" (May 25, 1997 to June 14, 1997)
at Villa Gualino, Torino.

\appendix

\section{Landau Levels on a torus}
 
Consider a magnetic field with potential 
${\bf A}=- B y $\^x.
The Hamiltonian is 
\begin{equation}
H={1\over 2m} [{(p_x+By)}^2+{(p_y)}^2] \, .
\label{3aaf}
\end{equation}
On a torus, we  identify
$z \sim z+m+n \tau$ with $\tau =\tau_1 +i\tau_2$ and $\tau_2 \geq 0$.   
The identification will impose boundary conditions 
on the wave function  
\begin{equation}
e^{it_x} \psi = e^{i\phi_1} \psi \, , e^{i\tau_1 t_x +i\tau_2 t_y}
\psi = e^{i\phi_2} \psi \, , 
\label{3aag}
\end{equation}
with
\begin{equation}
t_x=p_x \, , t_y=p_y+Bx 
\label{3aae}
\end{equation}
as magnetic translation operators and they commute with Hamiltonian.
The Dirac quantization condition $\tau_2 B=2\pi \phi$, with
the magnetic flux $\phi$  being an integer can be derived
by requiring operators $e^{it_x}, e^{i\tau_1 t_x +i\tau_2 t_y}$
commuting with each other for the consistence of  the boundary conditions
of eq.\ (\ref{3aag}). 

The wave function  describing an electron in the 
lowest Landau level has the form
\begin{equation}
\psi_l (x,y)=e^{-{ B y^2\over 2}}f(z) \, , 
\label{3aaa}
\end{equation}
where $f(z)$ is the holomorphic function.
Higher Landau levels can be obtained by acting
operator $a^+$ on $\psi_l$, $(a^+)^k\psi_l$,
with $a^+= \partial_{z} +{B(z-\bar z)\over 4}$.
We can write any  Landau level state as
\begin{equation}
\psi (x,y)=e^{-{ B y^2\over 2}}f(z,\bar z) .
\label{3aaahl}
\end{equation}
Now function $f$ is not necessary to be a holomorphic function
unless the wave function describes a lowest Landau level states.

By using the relation
\begin{equation}
e^{i\tau_1 t_x +i\tau_2 t_y}=e^{-i\tau_2 Bx^2 \over 2\tau_1}
e^{i\tau_1 p_x +i\tau_2 p_y}e^{i\tau_2 Bx^2 \over 2\tau_1}\, ,
\end{equation}
Eq.\ (\ref{3aag}) can be written as
\begin{equation}
f(z+1, \bar z +1)=e^{i\phi_1} f(z, \bar z) \, , 
f(z+\tau, \bar z +\bar \tau )
=e^{i\phi_2}e^{-i\pi \phi (2z+\tau)}f(z, \bar z) \, .
\label{3aah}
\end{equation}
In the case of lowest landau levels, 
$f$ is a holomorphic function, and the solutions of
eq.\ (\ref{3aah}) are theta functions.  
For the many-particle wave functions, 
the condition of  eq.\ (\ref{3aah}) shall 
be imposed on every particle.

\section{Theta functions}

$\theta$ function is defined as 
\begin{equation}
\theta (z|\tau ) =\sum _n \exp (\pi in^2 \tau +2\pi inz)\, ,  
n\subset integer \, . 
\label{3aai}
\end{equation}
We will generalize  the $\theta$ function  of eq.\ (\ref{3aai})  to the theta function 
on the lattice:
\begin{equation}
\theta (z|e, \tau ) =\sum _{n_i} \exp (\pi i v^2 \tau +2\pi iv\cdot z)
\, , 
\label{3aaj}
\end{equation}
where $v$ is a vector on a {\it l-dimension} lattice 
(we will call it $\Lambda$),  $v=\sum_{i=1}^l n_i e_i$, 
with $n_i$ being integers,  $e_i \cdot e_j =A_{ij}$ and 
$z$ is a vector on the lattice. 
$A_{ij}$ needs to be a positive definite matrix in order to 
have a well defined theta function.
The $\theta$ function in Eq.\ (\ref{3aai}) is a
special case of the $\theta$ function 
defined by Eq.\ (\ref{3aaj}) with 
$l=1, e_1 \cdot e_1 =1$.    
We define also
\begin{equation}
\theta {a\brack b} (z|e, \tau ) =\sum _{n_i} 
\exp (\pi i {(v+a)}^2 \tau +2\pi i(v+a)\cdot (z+b))\, , 
\label{3aak}
\end{equation}
where $a, b$ are  arbitrary constant vectors on the lattice. 
The dual lattice $e^{\ast}_i$ is defined as (we will call 
the dual lattice as $\Lambda^{\ast}$)
\begin{equation}
e_i^{\ast} \cdot e_j =\delta_{ij} \, , 
\label{3aal}
\end{equation}
then we have $e^{\ast}_i\cdot e^{\ast}_j=A^{-1}_{i,j}$.  
One can check that the following relations hold:
\begin{eqnarray}
& & \theta {a\brack b} (z+e_i|e, \tau )=e^{2\pi ia\cdot e_i}
\theta {a\brack b} (z|e, \tau )\, , \nonumber \\
& & \theta {a\brack b} 
(z+\tau e_i|e, \tau )=\exp {[-\pi i \tau e_i^2 -2\pi i e_i \cdot (z+b)]}
\theta {a\brack b} (z|e, \tau ) \, , \nonumber \\
& & \theta {a\brack b} (z+e^{\ast}_i|e, \tau )=e^{2\pi ia\cdot e^{\ast}_i}
\theta {a\brack b} (z|e, \tau )\, ,  \\
& & \theta {a\brack b} 
(z+\tau e^{\ast}_i|e, \tau )=\exp {[-\pi i \tau {(e_i^{\ast})}^2 
-2\pi i e^{\ast}_i \cdot (z+b)]}
\theta {a+e^{\ast}_i\brack b} (z|e, \tau ) \, , \nonumber
\label{3aakm}
\end{eqnarray}
and
\begin{equation}
\theta {a+e_i\brack b+e^{\ast}_j} (z|e, \tau )=\exp (2\pi i a \cdot 
e_j^{\ast}) \theta {a\brack b} (z|e, \tau ) \, . 
\label{3aar}
\end{equation}

In a {\it 1-dimension} lattice with 
$e_1 \cdot e_1 =1$, $a=b=1/2$,
the $\theta$ function
is denoted as
\begin{equation}
\theta_3(z|\tau)=\theta {{1\over 2}\brack {1\over 2}} 
(z|\tau )\, , 
\label{3aal1}
\end{equation}
is an odd function of $z$. And we have equations 
\begin{eqnarray}
& & \theta_3(z+1|\tau)=e^{\pi i}
\theta_3(z| \tau )\, , \nonumber \\
& & \theta_3 
(z+\tau | \tau )=\exp {[-\pi i \tau  -2\pi i \cdot (z+{1\over 2})]}
\theta_3(z|\tau ) \, . 
\label{3aam}
\end{eqnarray}

\section{Laughlin wave function on a torus}

Laughlin wave function on a torus was obtained in ref.\ \cite{torus}.
The wave function could be written in a more compact 
form \cite{torus1,torus2}.  we will follow those constructions
in refs.\ \cite{torus1,torus2}.

The Laughlin-Jastrow wave function on the torus at the filling $1\over m$
($m$ is an odd positive integer) can be written as
\begin{eqnarray}
& &\Psi (z_i)=\exp (-{\pi \phi \sum_i y^2_i \over 
\tau_2})F(z_i)\, , \nonumber \\
& & F(z_i)=\theta {a\brack b} (\sum_i z_ie|e,\tau)
\prod_{i<j} {[\theta_3(z_i-z_j|\tau)]}^m \, , 
\label{3aan} 
\end{eqnarray}
where $\theta$  function is on a 
{\it 1-dimension} lattice, $e^2=m$ , $i=1, 2
\ldots ,  N$ with $N$ being the number of the electrons 
and $a=a^{\ast}e^{\ast}, b=b^{\ast}e^{\ast}$. Thus 
\begin{eqnarray}
 & &F(z_i+1)={(-1)}^{N-1}e^{2\pi a^{\ast} } 
F(z_i)\, , \nonumber \\
& & F(z_i+\tau)=\exp (-\pi (N-1)-2\pi i b^{\ast})
\exp [-i\pi mN(2z_i+\tau)] F(z_i)\, .
\label{3aao}
\end{eqnarray}
Compared to Eq.\ (\ref{3aah}), we get 
\begin{equation}
\Phi=mN, \phi_1=\pi  (\phi +1)+2\pi n_1 +2\pi a^{\ast} , 
\phi_2=\pi (\phi +1)+2\pi n_2 -2\pi b^{\ast} \, . 
\label{3aap}
\end{equation}
Eq.\ (\ref{3aap}) has solutions
\begin{eqnarray}
& &a^{\ast} =a_0+i, \, \, ,  
b^{\ast}=b_0  \, \, , i=0,1,\ldots , m-1  \, , \nonumber \\
& & a_0={\phi_1 \over 2\pi }+{\phi +1\over 2} \, \, , 
b_0=-{\phi_2 \over 2\pi }+{\phi+1\over 2} \, , 
\label{3aas}
\end{eqnarray}
which  will give  $m$ orthogonal Laughlin-Jastrow 
wave function (other solutions are not independent on the solutions
given by Eq.\ (\ref{3aas}), which  
can be seen from Eq.\ (\ref{3aar})).
So there is a $m$-fold center-mass 
degeneracy.

\end{document}